\documentclass[%
 reprint,
superscriptaddress,
 amsmath,amssymb,
 aps,
 prl,
longbibliography
]{revtex4-1}

\usepackage{graphicx}
\usepackage{dcolumn}
\usepackage{bm}

\begin{document}

\title{Airy wave packets accelerating in space-time}

\author{H. Esat Kondakci}
\email{esat@creol.ucf.edu}
\affiliation{CREOL, The College of Optics \& Photonics, University of Central Florida, Orlando, Florida 32816, USA}%

\author{Ayman F. Abouraddy}%
\email{raddy@creol.ucf.edu}
\affiliation{CREOL, The College of Optics \& Photonics, University of Central Florida, Orlando, Florida 32816, USA}%

\begin{abstract}
Although diffractive spreading is an unavoidable feature of all wave phenomena, certain waveforms can attain propagation-invariance. A lesser-explored strategy for achieving optical selfsimilar propagation exploits the modification of the spatio-temporal field structure when observed in reference frames moving at relativistic speeds. For such an observer, it is predicted that the associated Lorentz boost can bring to a halt the axial dynamics of a wave packet of arbitrary profile. This phenomenon is particularly striking in the case of a self-accelerating beam -- such as an Airy beam -- whose peak normally undergoes a transverse displacement upon free-propagation. Here we synthesize an acceleration-free Airy wave packet that travels in a straight line by deforming its spatio-temporal spectrum to reproduce the impact of a Lorentz boost. The roles of the axial spatial coordinate and time are swapped, leading to `time-diffraction' manifested in self-acceleration observed in the propagating Airy wave-packet frame. 
\end{abstract}

\pacs{Valid PACS appear here}
\maketitle

\maketitle


Diffraction is a fundamental feature of all wave phenomena that leads to the spatial spreading of localized excitations upon free propagation. In optics, diffractive spreading sets the limit on imaging resolution, from microscopes to self-driving cars. Despite inexorable diffraction, there are particular waveforms that propagate self-similarly, and are thus called `diffraction-free'; e.g., Bessel beams \cite{Durnin1987}, among other possibilities \cite{Levy2016}. An altogether different approach -- that has received scant attention -- to hinder the diffraction of freely propagating fields \textit{without} recourse to specific waveforms relies on exploiting the impact of Lorentz transformations associated with observers moving with respect to the field \cite{Longhi2004, Saari2004}. In general, the characteristics of an optical field in space and time can vary drastically when observed in references frames moving at relativistic speeds. For example, the Lorentz boost associated with an observer moving along the propagation axis of a monochromatic beam reveals a wave packet whose spatio-temporal spectral components are correlated \cite{Longhi2004}. Furthermore, it has been recently predicted that novel and yet-to-be-observed phenomena take place after applying a Lorentz boost, such as the emergence of non-axial components of the orbital angular momentum and time-diffraction \cite{Bliokh2012}. Such experiments, however, offer prohibitive technical difficulties because of the relativistic speeds required.

Here we demonstrate that under a Lorentz boost the spatio-temporal (ST) profile of a pulsed beam (or wave packet) undergoes a spectral deformation that \textit{swaps} the roles of the axial spatial coordinate with that of time \cite{Longhi2004,PorrasArxiv17}, thereby bringing about two surprising consequences. The first is the total arrest of any axial dynamics and, secondly, `time-diffraction' is anticipated in the transverse plane; i.e., the usual spatial diffraction dynamics -- which is now halted -- is displayed instead in the local \textit{time} domain of the pulse. We confirm our predictions experimentally by realizing the ST spectral deformation associated with a Lorentz boost via a ST phase-only modulation of a pulsed laser beam \cite{KondakciArxiv17}.

Although these consequences apply to \textit{all} optical fields, such phenomena will be particularly striking and easily recognizable in the case of (1+1)D pulsed light sheets; that is, a wave packet described with one transverse dimension in addition to time. Light sheets pose a fundamental difficulty: no ‘diffraction-free’ monochromatic solutions exist -- with the exception of the Airy beam \cite{Berry1979} that indeed travels with no change in shape or size, but whose peak undergoes a transversal displacement upon axial propagation to trace out a parabolic trajectory \cite{Siviloglou2007}. These ‘self-accelerating’ beams \cite{Kaminer2012} have applications spanning microscopy \cite{Jia2014}, micro-particle manipulation \cite{Baumgartl2008}, and nonlinear optics \cite{Polynkin2009} by virtue of their salutary characteristics \cite{Broky2008, Hu2010, Efremidis2010}. We pose here the following question: How does the acceleration associated with an Airy wave packet appear to an observer moving at a relativistic speed? By deforming its ST spectral locus, we observe a halt in the usual transverse acceleration of an Airy wave packet, which propagates ‘acceleration-free’ with no changes in shape or scale along a \textit{straight} line. Furthermore, by reconstructing the complex field in space and time we observe a clear display of time-diffraction -- where ST acceleration is regained locally in the wave packet’s time frame.

\begin{figure*}[ht!]
	\includegraphics[scale=1.05]{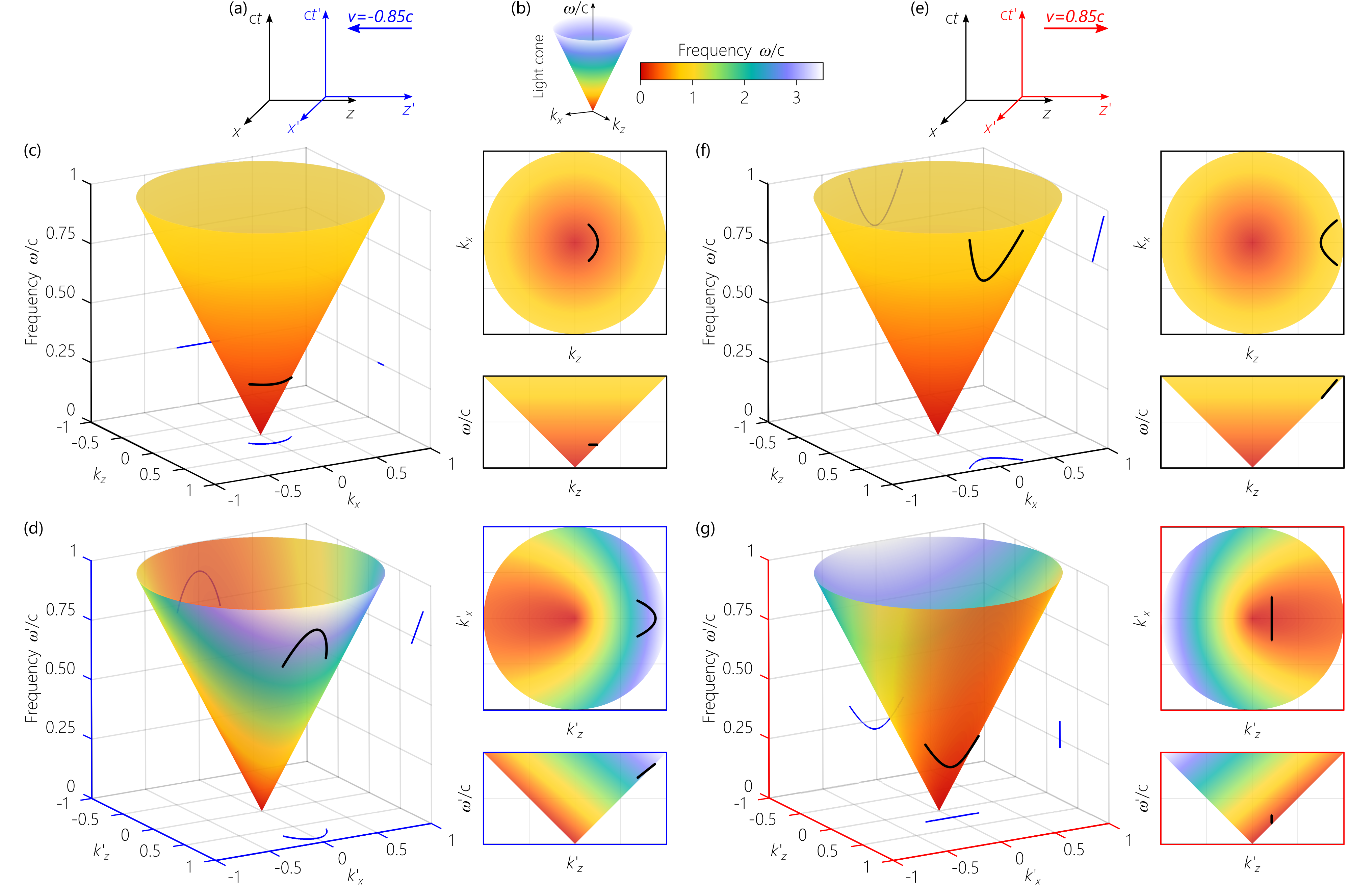}
	\caption{Deformation of the ST spectrum of an optical wave packet after implementing a relativistic Lorentz transformation. (a) The laboratory frame is described by the ST coordinates $(x,z,ct)$ and frequencies $(k_x,k_z,\omega/c)$. Primed coordinates are used for a frame moving at a velocity $v=-0.85c$ with respect to the laboratory frame along $z$. (b) The light-cone $\omega^2/c^2=k_x^2+k_z^2$,  where we use color for clarity to signify the frequency up to $\omega/c=3.5$ (in arbitrary units). (c) The spectral locus of a monochromatic beam in a horizontal plane $\mathcal{P}(0)$. (d) The monochromatic beam in (c) when viewed in the $(x',z',ct')$-frame becomes a wave packet along a conic section (here an ellipse) in the plane $\mathcal{P}(\theta)$; $\tan\theta=0.85=v_g/c$. Projections of this curve on the three planes $(k_x,k_z)$, $(k_x,\omega/c)$, and $(k_z,\omega/c)$ are shown. (e) The $(x',z',ct')$-frame is moving at a velocity $v=0.85c$. (f) The ST spectral locus of a wave packet in the $(x,z,ct)$-frame lying in a plane $\mathcal{P}(\theta)$ with $\tan\theta=1.176$. (g) The ST spectral locus of the wave packet in (f) observed in the moving $(x',z',ct')$-frame now lies in the $\mathcal{P}(\pi/2)$ plane. In the $(k_z',\omega'/c)$-plane, the projection is a vertical line. Note that the maximum frequency in (b,c) and (f,g) is $\omega/c=1$. The light-cone is preserved in the moving frames, but the frequencies in each constant-$k_x$ plane are red-shifted. We indicate this by retaining the same colors associated with the frequencies in (b).}
\label{fig:1} 
\end{figure*}

\begin{figure*}[!t]
	\includegraphics[scale=1.02]{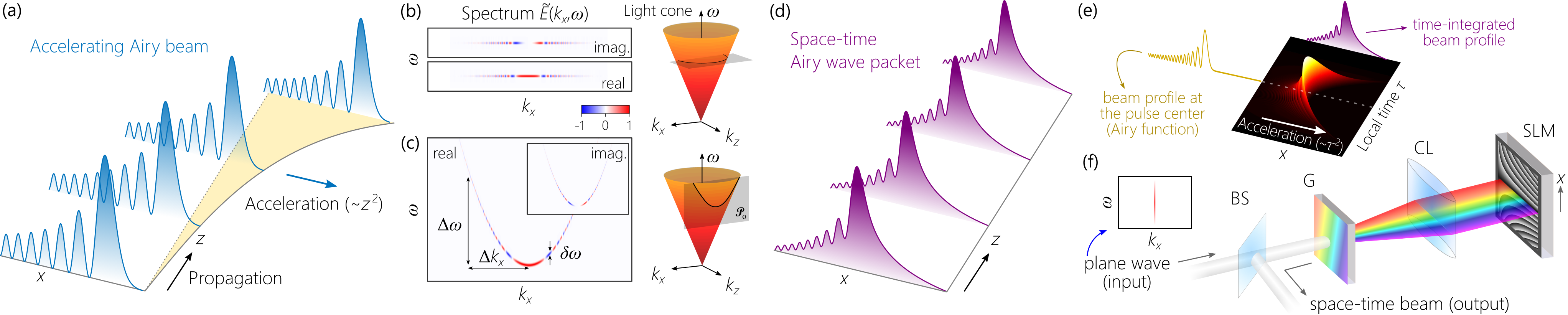}
	\caption{Airy wave packets accelerating in space-time. (a) Intensity of a traditional quasi-monochromatic Airy beam accelerating along a parabolic trajectory in the transverse direction x as it propagates along $z$. (b) The real and imaginary parts of the ST spectrum $\tilde{E}(k_x,\omega)$ are separable in $k_x$ and $\omega$ and has a cubic phase dependence on $k_x$. The ST spectrum approximately lies along the intersection of the light-cone with a horizontal iso-$\omega$ plane (inset). (c) Upon an appropriate Lorentz transformation, the ST spectrum $\tilde{E}(k_x,\omega)$ lies along the hyperbola at the intersection of the light-cone with an iso-$k_z$ plane. The real and imaginary parts are plotted highlighting the cubic phase with $k_x$ characteristic of an Airy beam along the correlated ST spectrum. (d) The ST Airy wave packet does not exhibit the expected acceleration and instead travels in a straight line. We plot here the time-averaged intensity $\int \mathrm{d}t|E(x,z,t)|^2$ as registered by a slow detector. (e) Acceleration is restricted to the local ST domain of the local time frame $\tau $ of the traveling pulse, when parabolic space-time curves are observed. At the pulse center $\tau =0$, the beam profile takes the precise shape of an Airy function. (f) Schematic of the setup for ST synthesis. BS: Beam splitter; CL: cylindrical lens; G: diffraction grating; SLM: spatial light modulator.}
\label{fig:2} 
\end{figure*}

We start by elucidating the impact of Lorentz boosts on (1+1)D optical fields of the form $E(x,z,t)=\psi(x,z,t)e^{i(k_o z-\omega_o t)}$, where $x$ and $z$ are the transverse and axial coordinates, respectively, $t$ is time, $\psi(x,0,t)$ is the envelope at $z=0$, $\omega_o$ is the carrier frequency, $k_o=\omega_o/c$ is a fixed wave number, and $c$ is the speed of light in vacuum [Fig.~\ref{fig:1}(a)]. The locus of the ST spectrum of any field in the laboratory frame $(x,z,ct)$ can be represented on the surface of the light-cone $k_x^2+k_z^2=(\omega /c)^2$, where $k_x$ and $k_z$ are the transverse and axial wave vector components (spatial frequencies), respectively, and $\omega$ is the angular (temporal) frequency [Fig.~\ref{fig:1}(b)]. The Lorentz boost associated with a reference frame $(x',z',ct')$ moving at a velocity $v$ with respect to the laboratory frame preserves the shape of the light-cone, but the points constituting its surface undergo a topology-preserving homeomorphism according to $k_x=k_x'$, $k_z=\gamma (k_z'+\beta\omega'/c)$, and $\omega/c=\gamma(\omega'/c+\beta k_z')$; here the Lorentz factor is $\gamma=1/\sqrt{(1-\beta ^2)}$, where $\beta =v/c$. In physical space, the Lorentz-transformed field is $E'(x',z',t')$ after substituting $x=x'$, $z=\gamma(z'+\beta ct')$, and $ct=\gamma(ct'+\beta z')$ in $E(x,z,t)$. Therefore, starting from a monochromatic beam $E(x,z,t)=\psi(x,z) e^{i(k_o z-\omega_o t)}$ whose ST locus is the circle at the intersection of the light-cone with a horizontal iso-$\omega$ plane $\mathcal{P}(0)$ [Fig.~\ref{fig:1}(c)], the Lorentz boost associated with the $(x',z',ct')$-frame deforms this spectral locus through tilting the iso-$\omega$ plane $\mathcal{P}(0)$ an angle $\tan\theta=-v/c$ with respect to the $(k_x,k_z)$-plane [Fig.~\ref{fig:1}(d)]. In other words, a \textit{monochromatic beam} in the laboratory frame becomes from the perspective of an observer in the $(x',z',t')$-frame a \textit{pulsed beam} or wave packet $E'(x',z',t')=\psi(x',z'-v_g t') e^{i(k_o' z'-\omega_o' t')}$; $\omega_o'=\gamma(1-\beta )\omega_o$ is a Doppler-shifted carrier frequency and $k_o'=\omega_o'/c$ \cite{Belanger1986}. The spectral locus of this wave packet is the conic section at the intersection of the tilted ST plane $\mathcal{P}(\theta)$ with the light cone, where $v_g=-v$ is the wave-packet group velocity by virtue of the linear dispersion relation in the $(k_z,\omega/c)$-plane, which is of the form $\omega/c=k_o+(k_z-k_o ) v_g/c$.

In general, the ST spectrum of a generic wave packet occupies a patch on the surface of the light-cone because of the inclusion of both ST frequency components whose complex amplitudes determine the spatial and temporal profiles. Wave packets lying in planes $\mathcal{P}(\theta)$ with $\theta\neq0$ are characterized by the property that each spatial frequency $k_x$ is correlated uniquely to a temporal frequency $\omega$, thus resulting in a reduced-dimensionality ST spectrum. Such wave packets undergo rigid free propagation along $z$ \cite{Belanger1986}. 

Starting instead with an ST wave packet of group velocity $v_g$,  $E(x,z,t)=\psi(x,z-v_g t) e^{i(k_o z-\omega_o t)}$ , a further Lorentz boost [Fig.~\ref{fig:1}(e)] results in additional tilting of the ST spectral plane that yields a wave packet of group velocity $v_g'$ through the usual relativistic addition of velocities $v_g'=(v_g-v)/(1-v_g v/c^2 )$. As such, no sequence of Lorentz boosts can produce a wave packet with $v_g>c$. Nevertheless, starting from a wave packet with $v_g>c$ lying in the plane $\mathcal{P}(\theta)$ with $\theta>\pi/4$ \cite{Wong2017, Sainte-Marie2017} as shown in Fig.~\ref{fig:1}(f), a Lorentz boost can produce a wave packet with arbitrary group velocity  $v_g'$ whose field is given by $E'(x',z',t' )=\psi(x',\gamma_z z'-\gamma_t ct')e^{(i(k_o' z'-\omega_o' t'))}$; where $\gamma_z=\gamma(1-\beta  \tan\theta )$, $\gamma_t=\gamma(\tan\theta-\beta )$, $v_g'=c\gamma_t/\gamma_z$, and the wave packet lies in a plane $\mathcal{P}(\theta')$ where $\tan\theta'=(\tan\theta-\beta )/(1-\beta  \tan\theta)$. This is not a violation of relativity, but instead an instance of the ‘scissors’ effect where the meeting point of the two blades can move at an arbitrary speed without transmitting information, as is the case of X-waves for instance \cite{Saari1997, Turunen2010, Hernandez2014}. The continuum of planes $\mathcal{P}(\theta)$ is thus segmented into two classes, corresponding to wave packets with subluminal and superluminal group velocities, each of whose members can be interconverted through Lorentz boosts. At the specific velocity $v=c^2/v_g=c/\tan\theta$ for the $(x',z',ct')$-frame, the Lorentz boost rotates $\mathcal{P}(\theta)$ such that it becomes \textit{exactly} the vertical iso-$k_z$ plane $\mathcal{P}(\pi/2)$ in Fig.~\ref{fig:1}(g) \cite{Longhi2004} , and the field simplifies to $E'(x',z',t' )=\psi(x',-\tan\theta ct'/\gamma)e^{(i(k_o z'-\omega_o t')/\gamma)}$. Wave packets lying in the plane $\mathcal{P}(\pi/2)$ are thus distinguished from those in any other plane $\mathcal{P}(\theta)$ that the $z$-dependence of the envelope of $E'(x',z',t')$ vanishes altogether and, consequently, all axial dynamics for such wave packets is halted except for an overall phase \cite{Kondakci2016b, Parker2016}. Furthermore, the \textit{same} ‘time-diffraction’ dynamics is observed in every transverse plane along the axis.

In specializing to an Airy wave packet, we first note that the unique monochromatic propagation-invariant solution to the wave equation in one transverse spatial dimension in the paraxial limit is of the form $E(x,z,t)=\psi(x-f(z))e^{i(\varphi (x,z)+k_o z-\omega_o t)}$ , where $\psi(x)=\mathrm{Ai}(x)$ is the Airy function, $f(z)$ traces a parabolic trajectory, and $\varphi $ is a phase factor \cite{Berry1979, Siviloglou2007}; Fig.~\ref{fig:2}(a). Such a beam can be readily synthesized in the spatial-frequency domain by recognizing that the Fourier transform of the Airy function is a cubic phase [Fig.~\ref{fig:2}(b)] \cite{Siviloglou2007}. A \textit{pulsed} Airy beam incorporates a finite spectral bandwidth such that $\psi(x,0,t)\approx g(t)\mathrm{Ai}(x)$, where $g(t)$ is the pulse profile. The reduced dimensionality envelope associated with the hyperbolic ST spectral locus in Fig.~\ref{fig:1}(b) once rotated to the iso-$k_z$ plane $\mathcal{P}(\pi/2)$ of the $(x',z',ct')$-frame takes the form $|\psi(x',z',t' )|=|\mathrm{Ai}(x'-f(-\tan\theta ct'/\gamma))|$, so that time-diffraction takes the form of acceleration along a parabolic trajectory in the ST domain $(x',t')$ independently of $z'$. For an Airy beam, this implies the elimination of transverse acceleration that is normally considered to be a fundamental aspect of its behavior.

\begin{figure*}[!t]
	\includegraphics[scale=1.4]{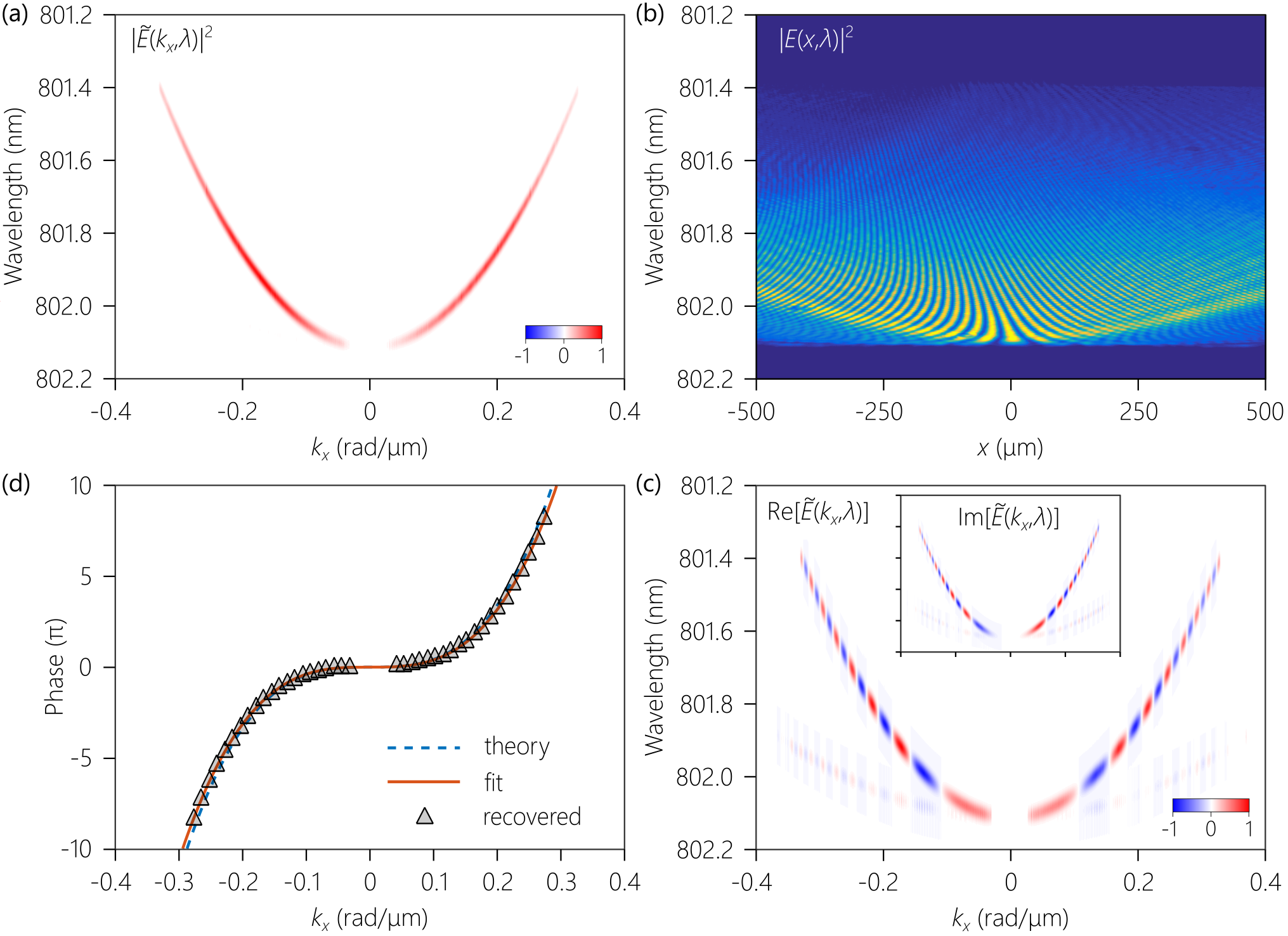}
	\caption{Retrieval of the complex spatio-temporal spectrum. (a) Measured ST spectrum $|\tilde{E}(\lambda ,k_x )|^2$; see Supplemental Material for measurement setup. (b) Measured hybrid-spectrum. The shift in the fringes with decreasing wavelength is a result of the relative phase difference between the oppositely signed but equal-amplitude spatial frequencies. (c) By taking a one-dimensional Fourier transform of (b) along the $x$-direction, we obtain a hyperbolic complex spectrum from which we extract the phase along the arms of the hyperbola, plotted here along $k_x$ and is in good agreement with the required cubic phase of the Airy spectrum. (d) The complex ST spectrum $\tilde{E}(\lambda ,k_x )$ obtained by superposing the experimentally measured spectrum $|\tilde{E}(\lambda ,k_x )|^2$ in (a) with the retrieved phase (c). }
\label{fig:3} 
\end{figure*}

Although theoretical studies have considered the impact of Lorentz boosts on the ST structure of wave packets \cite{Longhi2004, Saari2004, Belanger1986}, implementing such boosts via moving detectors is not practical at the velocities required, and their impact has thus not been verified experimentally to date. To test the predictions described above, we utilize a methodology that enables precise control over the spectral correlations underlying the ST structure of a wave packet. Specifically, we combine pulse shaping with beam modulation in the two-dimensional pulse shaper illustrated in Fig.~\ref{fig:2}(f) to introduce the ST spectral deformation associated with rotating the plane $\mathcal{P}$ to the vertical orientation in Fig.~\ref{fig:1}(g) \cite{KondakciArxiv17}. The phase-only modulation that sculpts the joint ST spectral distribution of the wave packet enables us to synthesize the target ST Airy wave packet [Fig.~\ref{fig:2}(c)] and thus realize the impact of the attendant Lorentz boost.

Starting from a femtosecond pulsed plane wave, a diffraction grating together with a cylindrical lens spreads the spectrum in space before impinging on a reflective spatial light modulator (SLM) that imparts a spatial phase distribution $\Phi(x,y)$ to the wave front. The reflected field returns to the grating whereupon the pulse is reconstituted \cite{KondakciArxiv17}; Fig.~\ref{fig:2}(f). The phase $\Phi$ is designed to achieve two goals simultaneously. First, the equal-amplitude but oppositely signed spatial frequencies $±k_x$ are assigned to one temporal frequency $\omega$ (within some spectral uncertainty $\delta \omega$) according to the hyperbolic dispersion relationship $k_o^2=\omega^2/c^2-k_x^2$, which results in confining the ST spectrum to the trajectory defined by the intersection of the light cone with the iso-$k_z$ plane $k_z=k_o$ (Fig.~\ref{fig:2}(c) inset). Second, a phase distribution of the form $e^{i\alpha k_x^3}$ is incorporated into $\Phi$ by the SLM to realize the functional form of the Airy beam. The reconstituted ST Airy wave packet is then expected to propagate in acceleration-free in a straight line.

To reconstruct the ST profile of the Airy wave-packet, we measure the amplitude and phase of the complex spectrum $\tilde{E}(k_x,\lambda )$. The magnitude $|\tilde{E}(k_x,\lambda )|$ is readily measured after implementing a spatial Fourier transform $x \rightarrow k_x$ upon the Airy wave packet using a $2f$ optical system \textit{and} resolving the spectrum for each $k_x$ via a diffraction grating (Supplemental Material). The measured squared-amplitude $|\tilde{E}(k_x,\lambda )|^2$ is shown in Fig.~\ref{fig:3}(a), which reveals a hyperbolic correlation between $k_x$ and $\lambda $, with temporal bandwidth $\Delta \lambda =0.715$ nm, spectral uncertainty $\delta \lambda =24$ pm, and spatial bandwidth $\Delta k_x=0.33$ rad/$\mu$m. We measure the phase of $\tilde{E}(k_x,\lambda )$ in two steps following the approach outlined in Ref. \cite{Dallaire2009}. First, we obtain the relative phase between the two branches of the hyperbola in Fig.~\ref{fig:3}(a) by measuring the spatially resolved spectrum $|\tilde{E}(x,\lambda )|^2$, which is plotted in Fig.~\ref{fig:3}(b). The sought-after phase is related to the shift in the fringes in $|\tilde{E}(x,\lambda )|^2$, which can be extracted by taking a Fourier transform of this hybrid spectrum along $x$. Second, we measure the spectral chirp in the original femtosecond pulse -- which is found to be negligible over the spectral bandwidth $\Delta \lambda $ -- utilizing a frequency-resolved optical gating (FROG)-based approach (Supplemental Material). The phase reconstructed by combining these two measurements reveals a cubic spectral phase that is in excellent agreement with the phase implemented by the SLM [Fig.~\ref{fig:3}(c)]. The complex ST spectrum $\tilde{E}(k_x,\lambda )$ resulting from combining the amplitude and phase measurements is given in Fig.~\ref{fig:3}(d).

\begin{figure}[!t]
	\includegraphics[scale=0.9]{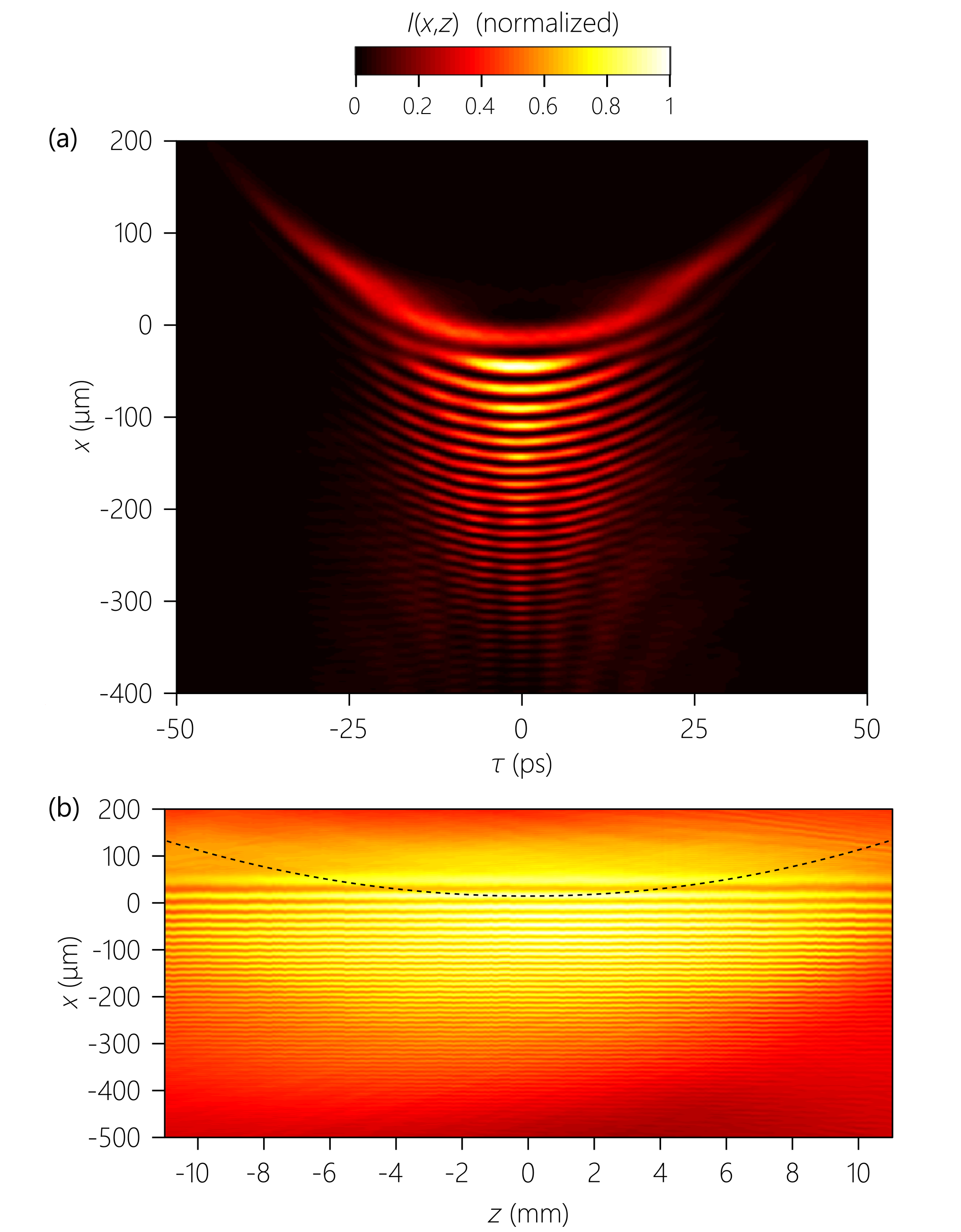}
	\caption{Transverse and axial measurements of a space-time Airy wave packet.  (a) Experimental spatio-temporal profile obtained by taking two-dimensional Fourier transform of the complex spectrum depicted in Fig.~\ref{fig:3}(d). (b) Intensity as a function of $x$ and $z$ detected by a CCD camera. The dashed line corresponds to the parabolic trajectory of a monochromatic Airy beam having the same beam waist.}
\label{fig:4} 
\end{figure}

It is now straightforward to obtain the complex ST field profile $E(x,\tau )$ of the Airy wave packet by performing a 2D ST Fourier transform $\tilde{E}(k_x,\lambda ) \rightarrow E(x,\tau )$, where $\tau $ is time in the moving wave packet frame; Fig.~\ref{fig:4}(a). It is clear that the pulse accelerates with $\tau $ in the $x$-direction along a local space-time trajectory given by $x=(\lambda _o c\tau )^2/(13.5\sigma_o )^3-2\sigma_o$, where $\sigma_o$ is the beam waist. The beam profile measured by a slow detector (e.g., a CCD camera) as a function of propagation distance $z$ is plotted in Fig.~\ref{fig:4}(b). The time-averaged wave-packet spatial profile does not exhibit any transverse dynamics as is usual with an Airy beam and instead propagates along a perfectly straight axial trajectory. A monochromatic Airy beam having the same spatial bandwidth $\Delta k_x$ would undergo a transverse shift of $\approx100$ $\mu$m along a parabolic trajectory over the course of $10$ mm axial propagation (the dashed black curve in Fig.~\ref{fig:4}(b)). Note that increasing the spectral uncertainty $\delta \lambda $, and thus reducing the tight ST correlations, reduces the acceleration-free range and deforms the Airy-function profile along $z$ (Supplemental Material).

We have produced the first acceleration-free Airy wave packet that propagates in a straight line. Our work contrasts with previous studies that have focused on Airy wave packets with separable spatial and temporal degrees of freedom any of which can take on the Airy functional form \cite{Abdollahpour2010, Chong2010, Valdmann2014}. An interesting avenue is the investigation of the interaction of these ultrafast diffraction-free wave packets with matter, especially electrons and other charged particles, in addition to excitations of exotic electronic states in 2D materials. Moreover, the concept of self-acceleration has ramifications that extend far beyond optical physics, extending from structuring electron beams \cite{Voloch-Bloch2013a}  to prolonging the lifetime of unstable relativistic fermions \cite{Kaminer2015}, all of which can benefit from the additional temporal degree of freedom introduced into the Airy wave packets in our work.

\begin{acknowledgments}
We thank D. N. Christodoulides and R. Menon for helpful discussions. This work was supported by the U.S. Office of Naval Research (ONR) under contract N00014-17-1-2458.
\end{acknowledgments}

\bibliography{library_short,books,arxiv}

\end{document}